\newcommand{\be}{\begin{equation}}\newcommand{\ee}{\end{equation}}
\newcommand{\bea}{\begin{eqnarray}}\newcommand{\eea}{\end{eqnarray}}
\newcommand{\nn}{\nonumber}\newcommand{\p}[1]{(\ref{#1})}
 \newcommand{\lb}[1]{\label{#1}}
 \newcommand\q{\quad}
\newcommand\qq{\quad\quad}

\def\a{\alpha}

\def\bt{{\beta}}

\def\g{\gamma}

\def\de{\delta}

\def\eps{\epsilon}

\def\ve{\varepsilon}

\def\vp{{\varphi}}

\def\la{\lambda}

\def\om{\omega}

 \newcommand\te{\theta}

\def\rh{\rho}

\def\z{\zeta}
\def\D{\Delta}

\def\G{\Gamma}

\def\La{\Lambda}
\def\Om{\Omega}
\def\T{\Theta}

\def\pa{\partial}

\newcommand\Tr{\mbox{Tr}\,}

\newcommand\ab{{\alpha\beta}}

\newcommand\cN{{\cal N}}

\newcommand\cM{{\cal M}}

\newcommand\cW{{\cal W}}

\def\sfrac#1#2{{\textstyle\frac{#1}{#2}}}

\documentclass[12pt]{article}
\usepackage{amsfonts}
\def\theequation{\arabic{section}.\arabic{equation}}

\begin{document}

\begin{center}
{\bf   THREE-DIMENSIONAL $\cN=4$ SUPERCONFORMAL\\ SUPERFIELD  THEORIES}\\

\vspace{0.5cm}
{\bf B.M. Zupnik}\\
\end{center}
{\it
Bogoliubov Laboratory of Theoretical Physics, JINR, Dubna,
  Moscow Region, 141980, Russia; E-mail: zupnik@theor.jinr.ru}
\begin{abstract}
The mirror map in the $D=3, \cN=4$ supersymmetry connects the left
and right $SU(2)$ automorphism groups and also the superfield
representations of the corresponding $\cN=4$ supermultiplets. The mirror
$\cN=4$ harmonic superspaces use the harmonics of two $SU(2)$ groups
and two types of the Grassmann analyticity. The irreducible left and right
$\cN=4$ supermultiplets are defined in these harmonic superspaces. We
analyze the $\cN=4$ superconformal interactions of the gauge and matter
superfields and the spontaneous breakdown of the superconformal
symmetry. The most interesting superconformal action possesses the
mirror symmetry and contains two nonlinear terms of the abelian left and
right gauge superfields, and also the mixing $\cN=4$ $BF$ interaction which
yields the topological masses of the gauge fields and the nontrivial
interaction of the scalar and pseudoscalar fields. The minimal interactions
of the left and right $\cN=4$ hypermultiplets can be included to this
abelian gauge theory.  We consider also the nonlinear $\cN=4$ gauge
superfield interactions.
\end{abstract}
Keywords: Harmonic superspace, extended supersymmetry,
superconformal symmetry \setcounter{equation}0
\section{Introduction}

Three-dimensional superconformal field theories with $\cN=6$ and
$\cN=8$ supersymmetries describe worldvolume of the $M2$-branes
at low energies \cite{BLG,BLS,ABJM}. The Lagrangians of these
theories contains matter the scalar and spinor fields interacting
with the Chern-Simons gauge vector fields for a specific choice
of the gauge groups. The three-dimensional $\cN=4$ Chern-Simons models
were studied in the $\cN=1$ superspace \cite{GW}. The algebras of
higher supersymmetries in these models close on the mass shell only.

The manifestly supersymmetric description of the three-dimensional
Chern-Simons-matter systems is possible for $\cN\leq 3$ supersymmetries
in the framework of the superspace approach \cite{Si},\cite{ZP1},
\cite{ZK},\cite{Z3}. These superfield constructions are universal for any
gauge groups. The $P$-parity is preserved in the gauge group $G\times G$
if we choose the difference of two  superfield Chern-Simons actions
corresponding to each $G$.

The $\cN=6$ Chern-Simons-matter model for the gauge group
$U(N)\times U(N)$ ($ABJM$-model) and the $\cN=8$ $BLG$-model for the
gauge group $SU(2)\times SU(2)$ were investigated in the $\cN=3$ harmonic
superspace \cite{BILPSZ}. The $\cN=3$ supersymmetry is manifest in this
formalism, the higher supersymmetry transformations connect different
superfields and the corresponding algebra of transformations closes on
the mass shell.

The superfield description of the $\cN=6$ Chern-Simons theory is possible
in the $SO(5)/U(2)$ or $SO(6)/U(3)$ harmonic superspaces \cite{HL},
\cite{Z8}. In the first version, the $\cN=6$ supersymmetry is realized on
the $\cN=5$ superfields. In the second version, the $SU(3)$ triplet of the
harmonic $\cN=6$ gauge superfields contains the Chern-Simons vector field
and the infinite number of auxiliary fields, for instance, the unusual
fermion field with three  spinor Lorentz indices and the $SO(6)$ indices.
All auxiliary fields vanish on the mass shell in these  variants
of the Chern-Simons theory, however, we do not know how to
include the  superfield matter interaction.

The $D=3, \cN=8$ Yang-Mills theory for the group $U(N)$ can be
constructed in the $\cN=4$ harmonic superspace by the analogy with the
four-dimensional case \cite{GIOS}, however, this theory  is not
superconformal in three dimensions.  In this paper, we analyze the
possible constructions of the $\cN=4$ superconformal  models in the
three-dimensional harmonic superspace \cite{Z3},\cite{Z5}.

The  $D=3, \cN=4$ superspace is covariant with respect to the Lorentz
group $SO(2,1)\sim SL(2,R)$ and the automorphism group
$SU_L(2)\times SU_R(2)$. The important property of the $\cN=4$
superspace is the discrete  symmetry with respect to the mirror map
\be
\cM:\hspace{1cm}SU_L(2)~\leftrightarrow~SU_R(2)\lb{Msym}
\ee
which connects the representations $(r,l)$ and $(l,r)$ of the group
$SU_L(2)\times SU_R(2)$. Different coordinate bases of the $\cN=4$
superspace are defined in appendix, for instance, the central basis
$(CB)$ is invariant under the $\cM$-map. The irreducible $D=3, \cN=4$
supermultiplets correspond to different superfield constraints, and
the mirror map connects the left and right versions of these constraints.
Respectively, we have the left and right hypermultiplets or the left and
right vector multiplets \cite{Z5}.

The left and right versions of the $\cN=4$ harmonic superspace use the
corresponding harmonics $u^\pm_k$ or $v^{(\pm)}_a$. In the appendix we
define the left analytic basis $(LAB)$ and the mirror right
analytic basis $(RAB)$ using left and right conditions of the
Grassmann analyticity. The left and right analytic superfields
describe the  mirror irreducible supermultiplets. In the next section,
we consider the superconformal transformations in different
representations of the $\cN=4$ superspace.

Section 3 is devoted to the superconformal interactions of the
$\cN=4$ gauge superfields. The left   abelian gauge prepotential is
defined as the analytic scalar superfield $V^{++}_0$ in the left
harmonic superspace. The corresponding pseudoscalar superfield
strength $W^{ab}$ satisfies the constraints of the right tensor
multiplet. The abelian superfield $W=\sqrt{W^{ab}W_{ab}}$ plays the
role of the dilaton superfield which helps us to construct the
superconformal actions in the $\cN=4$ superspace. We use $W$ as a
dynamical coupling constant in the superconformal version of the
$\cN=4$ abelian superfield gauge action $S^W_0$ \cite{Z5} and also
in the  superconformal interaction of the $\cN=4$ nonabelian gauge
superfields.

It is not difficult to obtain the right analytic gauge superfield
$A^{(++)}_0$ and the corresponding scalar left superfield strength
$L^{kl}$ using the mirror map. The mirror abelian superconformal action
 $S^L_0(A^{(++)}_0)$ contains the second dilaton $L=\sqrt{L^{kl}L_{kl}}$.

The superconformal interactions of the left hypermultiplet and the
improved tensor multiplet are studied in Sec. 4 by the analogy with the
corresponding interactions of the $D=4, \cN=2$ supermultiplets. The left
$D=3, \cN=4$ hypermultiplet has the minimal interaction only with the
left vector multiplet. It is shown that the superconformal abelian action
$S^L_0$ is equivalent to the analytic action of the improved tensor
multiplet, which is dual to the free action of the left hypermultiplet.

In Sec. 5 we consider the $\cN=4$ superfield terms generalizing the
Dirac-Born-Infeld  interactions with the derivatives of the vector and
scalar fields. These terms are invariant under the nonlinear
transformations of the $\cN=8$ supersymmetry.

We analyze the $\cN=4$ superconformal abelian  $BF$ term $S^0_{BF}$
which connects the left $U_L(1)$  scalar gauge superfield $V^{++}_{0}$
and and the right $U_R(1)$  pseudoscalar gauge superfield $A^{(++)}_0$
\cite{Z6}. This term was considered earlier in components \cite{BG},
\cite{KS}.  Note that the $\cN=4$ action $S^0_{BF}$ can be rewritten as
the difference of two abelian Chern-Simons terms in the $\cN=3$
superfield formalism \cite{BILPSZ}, in this case  the  transformations of
parity and the fourth supersymmetry connect the mirror abelian gauge
superfields defined in the single $\cN=3$ analytic superspace.

It is not possible to present the action $S^0_{BF}$ as the difference
of two  actions in the $\cN=4$ superspace, so we consider it as the
$\cM$-symmetric analog of the Chern-Simons action for the group
$U_L(1)\times U_R(1)$. The independent interactions of the mirror
$\cN=4$ hypermultiplets and the corresponding gauge multiplets can be
easily included to this picture. The  term $S^0_{BF}$ yields the specific
superconformal interaction of the left and right $\cN=4$ abelian gauge
superfields in the composite $\cM$-symmetric action $S^W_0+S^L_0+
S^0_{BF}$. This action describes the nontrivial interactions of the left
pseudoscalar and right scalar fields with the topologically massive vector
and pseudovector fields in the bosonic sector. It is not possible to
formulate  the non-abelian Chern-Simons  models in the $D=3, \cN=4$
superspace.

\setcounter{equation}0
\section{ Superconformal $D=3, \cN=4$ transformations}
The $CB$ coordinates of the $\cN=4$ superspace $z^M=(x^m, \te^\a_{ak})$
are defined in appendix. The infinitesimal $\cN=4$ superconformal
transformations of these coordinates have the form
\bea
&&\de x^\ab=\la^\ab=c^\ab+a^\a_\g x^{\g\bt}+a^\bt_\g x^{\a\g}+b\,x^\ab
-i\eps^\a_{ka}\te^\bt_{ka}-i\eps^\bt_{ka}\te^\a_{ka}
+\sfrac12x^{\a\g}k_{\g\rh}x^{\rh\bt}+\sfrac18\T^2 k^\ab\nn\\
&&+\sfrac14x^{\a\rh}\T k^\bt_\rh-\sfrac{i}2\eta_{ka\g}\te^{ka\a}x^{\g\bt}
-\sfrac{i}2\eta_{ka\g}\te^{ka\bt}x^{\g\a}+
\sfrac14\eta^\bt_{ka}\T\te^{ka\a}+\sfrac14\eta^\a_{ka}\T\te^{ka\bt},\nn\\
&&\de\te^\a_{ka}=\la^\a_{ka}=\eps^\a_{ka}+a^\a_\bt\te^\bt_{ka}
+\sfrac12b\,\te^\a_{ka}-\om^l_k\te^\a_{la}-\Om^b_a\te^\a_{kb}+
\sfrac12x^\ab\te^\g_{ka}k_{\bt\g}\nn\\
&&-\sfrac{i}4\T\te^\g_{ka}k^\a_\g+\sfrac12x^\ab\eta_{{ka}\bt}
-i\te^\a_{lb}\te^\g_{ka}\eta^{lb}_\g-\sfrac{i}4\T\eta^\a_{ka},\lb{SC2}
\eea
where  $c^m, a^m, k_m, b$ are parameters of the conformal group
$SO(3,2)$, $\om_{kl}$ and $\Om_{ab}$ are the parameters of the group
$SU_L(2)\times SU_R(2)$, $\eps^\a_{ka}$ and $\eta^\a_{ka}$ describe the
$Q$ and $S$ supersymmetries, and $\T=\te^\g_{ka}\te^{ka}_\g$.

The standard notation of this superconformal group is $OSp(4|4)$, but we
sometimes use the short notation $SC$. The superconformal transformation
of the full-superspace integral measure $\de_{sc}d^{11}z=j(z)\,d^{11}z$
contains the superfield parameter
\bea
&&j(z)=\pa_m\la^m-\pa_\a^{ia}\la^\a_{ia}=-b-k_mx^m
-i\te^{la\bt}\eta_{la\bt}.\lb{SC3}
\eea

The superconformal transformations of the flat vector  differential
$\om^\ab=dx^\ab-id\te^\a_{ka}\te^{ka\bt}-id\te^\bt_{ka}\te^{ka\a}$ and the
spinor derivative $D^{ka}_\a$ have the covariant form
\bea
&&\de_{sc}\om^\ab=-j\om^\ab+\chi^\a_\rh\om^{\rh\bt}+\chi^\bt_\rh\om^{\a\rh},
\nn\\
&&\de_{sc}D^{ka}_\a=\sfrac12jD^{ka}_\a-\chi^\rh_\a D^{ka}_\rh+
\la^k_lD^{la}_\a+\xi^a_c D^{kc}_\a.\lb{SCD} \eea
Three traceless $2\times 2$ transformation matrices are constructed from
the parameters of $OSp(4|4)$
\bea
&&\chi^\rh_\bt=\sfrac12a^\rh_\bt+\sfrac12(x^{\a\la}k_{\bt\la}
-\sfrac12\de^\rh_\bt x^{\g\la}k_{\g\la})-\sfrac{i}4\te_{ka}^\la\te^{ka}_\la
k^\rh_\bt+i\te_{ka}^\rh\eta^{ka}_\bt-
\sfrac{i}2\de^\rh_\bt\te_{ka}^\la\eta^{ka}_\la,\nn\\
&&\xi^{bc}=\Om^{bc} -\sfrac{i}2\te^{lb\bt}\te^{c\g}_l k_{\bt\g}
-\sfrac{i}2(\te^{lb\g}\eta^c_{l\g}+\te^{lc\g}\eta^b_{l\g}),\\
&&\la^{kl}=\cM \xi^{bc}=\om^{kl}-\sfrac{i}2\te^{kc\bt}\te^{l\g}_c k_{\bt\g}
-\sfrac{i}2(\te^{kc\g}\eta^l_{c\g}+\te^{lc\g}\eta^k_{c\g})\nn
\eea
and satisfy the simple relations
\bea
&&D^{ka}_\g\chi^\rh_\bt=-\de^\rh_\g D^{ka}_\bt j+\sfrac12\de^\rh_\bt
D^{ka}_\g j,\q D^{ka}_\g\xi^{bc}=\sfrac12\ve^{ab}D^{kc}_\g j+
\sfrac12\ve^{ac}D^{kb}_\g j,\nn\\
&&D^{nb}_\a \la^{kl}=\sfrac12\ve^{nk}D^{lb}_\a j+\sfrac12\ve^{nl}D^{kb}_\a j
,\q D^{lb}_\g j(z)=-i\te^{lb\la}k_{\g\la}-i\eta^{lb}_\g.\lb{SC10}
\eea

By analogy with \cite{GIOS}, we define the following $OSp(4|4)$
transformations of the left harmonics:
\bea
&&\de_{sc} u^+_k=\la^{++}u^-_k,\q \de_{sc}u^-_k=0,\q \nn\\
&&\la^{++}=u^+_ju^+_l\la^{jl}(z)=\om^{++}-\sfrac{i}2\te^{+a\a}\te^{+\bt}_a
k_\ab-i\te^{+a\a}\eta^+_{a\a}. \lb{SC4}
\eea
The $SC$ transformations of the  coordinates $x^m_L, \te^{+\a}_a$ in $LAB$
\p{LAB} are manifestly analytic
\bea
&&\la^m_L=\de_{sc}x^m_L=\sfrac12(\g^m)_\ab \de_{sc}x^\ab_L= b\,x^m_L
+(x_Lk)x^m_L-\sfrac12x^2_Lk^m\nn\\
&&-2i(\g^m)_\ab\eps^{-\a}_{b}\te^{+b\bt}
-i(\g^m)_\ab(\g^n)^{\rh\bt}\eta^-_{b\rh}\te^{+b\a}x_{nL}
-i(\g^m)_\ab\om^{--}\te^{+\a}_b\te^{+b\bt},\\
&&\la^{+\a}_{a}=\de_{sc}\te^{+\a}_a=\eps^{+\a}_a+\sfrac12b\,\te^{+\a}_a
+\om^{+-}\te^{+\a}_a-\Om^b_a\te^{+\a}_b+\sfrac12x^\ab_L\te^{+\g}_ak_{\bt\g}
\nn\\
&&-i\te^{+\g}_a\te^{+\a}_b\eta^{-b}_\g
\eea
where the standard Poincare transformations are omitted. We define the
$SC$ transformation of the left analytic integral measure
\bea
&&\de_{sc}d\z^{-4}du=(\pa^L_m\la^m_L+\pa^{--}\la^{++}-\pa^{-a}_\a\la^{+\a}_a)
d\z^{-4}du=-2\la d\z^{-4}du,\nn\\
&&\la=\om^{+-}-\sfrac12(b+k_mx^m_L)+i\te^{+\a}_a\eta^{-a}_\a,\q
D^{++}\la=\la^{++}.\lb{SC5}
\eea

The superconformal parameters in $CB$ \p{SC3} and $LAB$ are connected by
the simple relations
\be
j(z)=2\la-D^{--}\la^{++},\q \la=\sfrac12j+u^+_ju^-_n\la^{jn}(z).\lb{CBL}
\ee

We consider the superconformal transformation of the non-analytic spinor
coordinates
\bea
&&\la^{-\a}_{a}=\de_{sc}\te^{-\a}_a=\eps^{-\a}_a+\sfrac12b\te^{-\a}_a
+a^\a_\bt\te^{-\bt}_a-\Om^b_a\te^{-\a}_b-\om^{+-}\te^{-\a}_a+
\om^{--}\te^{+\a}_a+\sfrac12x^\ab_L\te^{-\g}_ak_{\bt\g}\nn\\
&&+i\te^{-\g}_a\te^{-\a}_b\te^{+b\bt}k_{\bt\g}+\sfrac12x^\ab_L\eta^-_{a\bt}
+i\te^{-\g}_a\te^{-\a}_b\eta^{+b}_\g+i\te^{-\a}_b\te^{+b\bt}\eta^-_{a\bt}
-i\te^{-\g}_a\te^{+\a}_b\eta^{-b}_\g.\lb{SC6}
\eea
This formula yields the $OSp(4|4)$ transformation of the
spinor derivative $D^{+b}_\bt$
\bea
&&\de_{sc}D^{+b}_\bt=-(D^{+b}_\bt \de_{sc}\te^{-\rh}_c)D^{+c}_\rh=
-\chi_\bt^\rh D^{+b}_\rh+\xi^b_c D^{+c}_\bt+\la D^{+b}_\bt\lb{SCD+}
\eea
where the $LAB$ superconformal matrices $\chi_\bt^\rh$ and $\xi^b_c$
are identical to the corresponding matrices in $CB$ \p{SCD}.

We obtain the $SC$ transformations of the  higher spinor derivatives
from this formula
\bea
&\de_{sc}D^{++}_\ab=-\chi_\a^\rh D^{++}_{\rh\bt}-\chi_\bt^\rh D^{++}_{\rh\a}
+2\la D^{++}_\ab+(D^{+b}_\a j)D^+_{b\bt}+(D^{+b}_\bt j)D^+_{b\a},&\nn\\
&\de_{sc}D^{++ab}=-(D^{+a\rh}j)D^{+b}_\rh-(D^{+b\rh}j)D^{+a}_\rh
+\xi^a_cD^{++cb}+\xi^b_cD^{++ac}+2\la D^{++ab},\lb{SC9}&\\
&\de_{sc}(D^+)^4=4\la (D^+)^4.&\nn
\eea

The superconformal transformations of the left harmonic derivatives are
\bea
&&\de_{sc}D^{++}=-\la^{++}D^0,\q \de_{sc}D^{--}=-(D^{--}\la^{++})D^{--}.
\lb{SC8}
\eea

The $SC$ transformations in the right harmonic superspace can be
obtained by the mirror map  from the $LAB$ transformations, for
instance,
\bea
&&\de_{sc} v^{(+)}_a=\xi^{(++)}v^{(-)}_a,\q \de_{sc}v^{(-)}_a=0,\q
\de_{sc}\te^{(\pm)\a}_k=\cM \de_{sc}\te^{\pm\a}_a,
\q \de_{sc}x^m_R=\cM \de_{sc}x^m_L,\nn\\
&&\xi^{(++)}=v^{(+)}_av^{(+)}_b\xi^{ab}=v^{(+)}_av^{(+)}_b\Om^{ab}
-\sfrac{i}2\te^{(+)l\a}\te^{(+)\bt}_l
k_\ab-i\te^{(+)l\a}\eta^{(+)}_{l\a}.\lb{SC11}
\eea

\setcounter{equation}0
\section{Superconformal $\cN=4$ superfield gauge interactions}

We consider the  superfield constraints of the left hypermultiplet
$q^{ka}$ and the left tensor multiplet $L^{lj}=L^{jl}$ and their $SC$
transformations in $CB$
\bea
&&D^{ka}_\a q^{lb}+D^{la}_\a q^{kb}=0,\q \de_{sc}q^{lb}=\sfrac12j q^{lb}
+\la^k_n q^{nb}+\xi^b_cq^{lc},\lb{Lq}\\
&&D^{ka}_\a L^{jl}+D^{ja}_\a L^{lk}+D^{la}_\a L^{kj}=0,
\q \de_{sc}L^{kl}=jL^{kl}+\la^k_nL^{nl}+\la^l_nL^{kn}.\lb{Lte}
\eea
The mirror map \p{Msym} allow us to obtain the superfield constraints
for the right hypermultiplet $Q^{ka}$ and the right tensor multiplet $W^{ab}$
\bea
&&D^{ka}_\a Q^{lb}+D^{kb}_\a Q^{la}=0,\lb{Rq}\\
&&D^{ka}_\a W^{bc}+D^{kb}_\a W^{ca}+D^{kc}_\a W^{ab}=0.\lb{Rte}
\eea
The superfield $W^{ab}$ can be treated as the superfield strength of the
left abelian spinor gauge superfield, and the mirror superfield $L^{kl}$
has a similar interpretation. $\cM$-map can be combined with the
$P$-parity, then the mirror superfields $W^{ab}$ and $L^{kl}$ have
opposite parities.

In the $\cN=4$ harmonic superspace, the left non-abelian gauge
supermultiplet is described by the matrix analytic prepotential
$V^{++}(\z_L,u)$
\bea
&&\de_\La V^{++}=-D^{++}\La-[V^{++},\La],\q(V^{++})^\dagger=-V^{++},\q
\Tr V^{++}=0, \nn\\
&&\La^\dagger=-\La,\q\Tr\La=0\lb{SUn}
\eea
where $\La(\z_L,u)$ is the analytic superfield matrix  parameter of the
gauge group $SU(N)$. We analyze the off-shell component fields
$\phi_{ab}, A_m, \la^{ka}_{\a}$ and $X^{kl}$ in the $WZ$-gauge of the
prepotential
\bea
&&V^{++}_{WZ}=\te^{+a\a}\te^{+b}_\a \phi_{ab}+\T^{++m}A_m
+2i\T^{+3\a}_au^-_k\la^{ka}_{\a}+3i(\te^+)^4u^-_ku^-_lX^{kl}\lb{WZ}
\eea
where $\T^{+3\a}_a$ and $(\te^+)^4$ are defined in appendix. The component
fields of the $P$-even superfield $V^{++}$ include the pseudoscalar
$\phi^{ab}$, spinor $\la^{ka}_\a$, vector $A_m$ and  auxiliary
scalar $X^{kl}$.

The non-analytic harmonic connection $V^{--}$ can be constructed in
terms of the prepotential \cite{Z2}
\be
V^{--}(z,u)=\sum_{n=1}^\infty (-1)^n \int du_1\ldots
du_n \frac{V^{++}(z,u_1)V^{++}(z,u_2)\ldots
V^{++}(z,u_n)}{(u^+u^+_1)(u^+_1u^+_2)\ldots (u^+_n u^+)}
\label{Vm}
\ee
where the harmonic distributions $(u^+_1u^+_2)^{-1}$\cite{GIOS} are used.

The $LAB$ representation of the nonabelian gauge superfield strength is
\cite{Z3,Z5}
\bea
&&\cW^{ab}=-\sfrac14D^{+a\a}D^{+b}_\a V^{--},\q
\de_\La \cW^{ab}=[\La,\cW^{ab}],\nn\\
&&D^{++}\cW^{ab}+[V^{++},\cW^{ab}]=0.\lb{Wstr}
\eea

The abelian $\cN=4$ superfield action contains the imaginary
prepotential $V^{++}_0$ and the coupling constant $g_3$ of
dimension $\sfrac12$
\bea
&&S^L_{QED}=-\frac1{4g_3^2}\int d^{11}z du V^{++}_0V^{--}_0\nn\\
&&=\frac1{4g_3^2}\int d^3x[-\phi^{ab}\Box \phi_{ab}+
2A^m\pa^nF_{nm}-2i\la^\a_{ka}\pa_{\ab}\la^{ka\bt}
+X^{kl}X_{kl}]\lb{QED}
\eea
where $F_{nm}=\pa_nA_m-\pa_nA_m$.

We consider the passive form of the superconformal transformations of
the harmonic connections \cite{GIOS}
\bea
\de_{sc}V^{++}=0,\q \de_{sc}V^{--}=-(D^{--}\la^{++})V^{--}\lb{Vsc}
\eea
or the active (local) form of the same transformations
\be
\de^*_{sc}V^{\pm\pm}(z,u)=-(\la^m\pa_m+\la_{ka}^\a\pa^{ka}_\a+\la^{++}D^{--})
V^{\pm\pm}(z,u)+\de_{sc}V^{\pm\pm},\lb{Vsc1}
\ee
where functions $\la^M$ and $\la^{++}$ are defined in the previous
section. The action $S_{QED}^L$ is not invariant under these
transformations.

The abelian superfield strength has the simple form
\bea
&&W^{ab}(V)
=-\frac14\int du D^{-a\a}D^{-b}_\a V^{++}_0(z,u),\nn\\
&&\de_\La W^{ab}=0,\q \overline{W^{ab}}=W_{ab}.\lb{Wstr2}
\eea
The superfield  $W^{ab}(V)$ is $P$-odd if the corresponding
gauge superfield $V^{++}$ is $P$-even. We consider its decomposition
in the left coordinates
\bea
&&W^{ab}(x_L,\te^{\pm}_a,u)=\phi^{ab}-i\te^{-a\bt}\te^{+\a}_{c}\pa_\ab^L
\phi^{bc}-i\te^{-b\bt}\te^{+\a}_{c}\pa_\ab^L \phi^{ac}
+\T^{--ab}\T^{++}_{cd}\Box\phi^{cd}\nn\\
&&+\frac{i}2(\te^{-b\a}\te^{+a\bt}+\te^{-a\a}\te^{+b\bt})F_\ab
+\frac{i}2(\te^{+a\a}u^-_k-\te^{-a\a}u^+_k)\la^{kb}_\a+\frac{i}2
(\te^{+b\a}u^-_k-\te^{-b\a}u^+_k)\la^{ka}_\a\nn\\
&&+i\te^{+a\g}\te^{+b}_\g u^-_ku^-_lX^{kl}+i\te^{-a\g}\te^{-b}_\g u^+_ku^+_lX^{kl}
-i\te^{+a\g}\te^{-b}_\g u^+_ku^-_lX^{kl}-i\te^{-a\g}\te^{+b}_\g u^+_ku^-_lX^{kl}
\nn\\
&&+\mbox{higher derivative terms}
\eea
where $F_\ab=\pa^L_{\a\g} A_\bt^\g+\pa^L_{\bt\g} A_\a^\g$.

The superfield $W^{ab}$ can be connected with the right analytic
superfield $W^{(++)}(\z_R,v)=v^{(+)}_av^{(+)}_bW^{ab}$. The component
decomposition of this  superfield representation is very short
in the right analytic coordinates \p{RAB}.

The $SC$ transformation of the nonabelian superfield strength can be
obtained using \p{Vsc} and \p{SC9}
\bea
&\de_{sc}\cW^{ab}=-\sfrac14\de_{sc}(D^{++ab}V^{--})=
j\cW^{ab}+\xi^a_c\cW^{cb}+\xi^b_c\cW^{ac}.&\lb{Wsc}
\eea

The important $SU_R(2)$ invariant abelian dilaton superfield
\be
W=\sqrt{W^{ab}W_{ab}},\q \de_{sc}W=j(z)W\lb{W0}
\ee
satisfies the superfield constraint $D^{++}_{\ab}(W^{-1})=0$.

The superfield $W$ is used in the superconformal abelian gauge action
\cite{Z5}
\be
S^W_0(V)=-\frac14\int d^{11}z\,du\frac{1}{W}
V^{++}_0(z,u)V^{--}_0(z,u).\lb{S0}
\ee
The $SC$ invariance of $S^W_0$ can be checked straightforwardly using
transformations of the integral measure and all superfields.

The spontaneous breakdown of the superconformal symmetry arises if we
redefine the pseudoscalar  superfield $W^{ab}$
\bea
&&W^{ab}=\vp (C^{ab}+w^{ab}),\q C^{ab}C_{ab}=2,\nn\\
&&W=\vp\sqrt{2+2C^{ab}w_{ab}+w^{ab}w_{ab}}
\eea
where $w_{ab}$ is the improved superfield strength, $\vp$ is the scale
fixing parameter of dimension 1 and $C^{ab}$ are  dimensionless
constants  describing the spontaneous breakdown of the parity and
$SU_R(2)$ symmetry.

The abelian dilaton superfield $W(V^{++}_0)$ plays the role of
a dynamic coupling constant in the superconformal version of the
nonabelian gauge theory
\bea
S_N(W,V^{++})=\int d^{11}zdu_1\ldots du_n
\sum_{n=2}^\infty \frac{(-1)^{n+1}}{4Wn}  \frac{\Tr V^{++}(z,u_1)\ldots
V^{++}(z,u_n)}{(u^+_1u^+_2)\ldots (u^+_n u^+_1)}.\lb{SCN}&&
\eea
The $SC$ invariance of this action  can be checked by the method of
\cite{GIOS} using the active superconformal transformations
\bea
&&\de^*_{sc}V^{++}(z,u)=-(\la^M\pa_M+\la^{++}D^{--})
V^{++}(z,u),\nn\\
&&\de^*_{sc}(W^{-1})=(-\la^M\pa_M+D^{--}\la^{++}-2\la)W^{-1}.
\eea

The superfield constraint $W=g^2_3$ in \p{SCN} breaks down the conformal
symmetry, in this case we obtain the superfield action of the $\cN=4$
Yang-Mills theory $S_N(g_3^2,V^{++})$ constructed by analogy with the
four-dimensional $\cN=2$ gauge action \cite{Z2}.

The  right gauge multiplets can be considered as the mirror constructions
in the right harmonic superspace \p{RAB}. The abelian right analytic
prepotential $V^{(++)}_{R0}(\z_R,v)\equiv A^{(++)}_0(\z_R,v)$
has the gauge transformation
\be
 \de_\La A^{(++)}_0=-D^{(++)}\La_R(\z_R,v),\q D^{(+)k}_\a A^{(++)}_0=0.
\ee
In the $WZ$-gauge, this prepotential contains the component fields
of the right vector multiplet
\bea
&&A^{(++)}_{WZ}=\te^{(+)\a}_k\te^{(+)}_{l\a} \La^{kl}
+(\te^{(+)k}\g^m\te^{(+)}_k)B_m+\sfrac{4i}3\te^{(+)\bt}_k\te^{(+)}_{l\bt}
\te^{(+)l\a}v^{(-)}_a\rh^{ka}_{\a}\nn\\
&&+3i(\te^{(+)})^4v^{(-)}_av^{(-)}_bY^{ab}.\lb{WZR}
\eea
Below we use the pseudoscalar right prepotential $A^{(++)}_0$,
then $B_m$ is the pseudovector field, $\La^{kl}$ is the scalar field and
$Y^{ab}$ is the pseudoscalar auxiliary field.

We can construct the  superfield strength of the right abelian
prepotential
\bea
&&L^{kl}=-\frac14\int dv D^{(--)kl} A^{(++)}_0(z,v).\lb{Lkl}
\eea
satisfying the constraints of the left tensor multiplet \p{Lte}.
It is scalar for the $P$-odd prepotential $A^{(++)}_0$. The
superconformal transformation of the superfield $L^{kl}$
can be obtained by the mirror map from \p{Wsc}
\bea
\de_{sc}L^{kl}=j(z)L^{kl}+\la^k_j(z)L^{jl}+\la^l_j(z)L^{kj}.
\eea

The improved left superfield $l_{kl}$ is defined by the formula
\bea
&&L^{kl}=\g(c^{kl}+l^{kl}),\q c^2=c^{kl}c_{kl}=2
\eea
where $\g$ is some constant of dimension 1 and $c_{kl}$ are
dimensionless constants of the spontaneous breakdown of $SU_L(2)$.
The mirror dilaton superfield is
\be
L=\sqrt{L^{kl}L_{kl}}=\g\sqrt{2+2c_{kl}l^{kl}+l_{kl}l^{kl}},\q
\de_{sc}L=j L.\lb{DL}
\ee
The mirror superconformal abelian interaction contains this  superfield
$L$
\be
S^L_0=-\frac14\int d^{11}z\,dv\frac{1}{L} A^{(++)}_0A^{(--)}_0,\q
D^{(++)}A^{(--)}_0=D^{(--)}A^{(++)}_0.\lb{SL0}
\ee
In Sec. 4 we analyze the equivalent left analytic action of the
improved tensor multiplet which is dual to the free hypermultiplet
action.

We use the spinor derivative of the superfield $L$ in the
superconformal spinor connection
\bea
\G^{ka}_\a=L^{-1}D^{ka}_\a L,\q \de_{sc}\G^{ka}_\a= D^{ka}_\a j+
\sfrac12j \G^{ka}_\a-\chi_\a^\rh \G^{ka}_\rh
+\la^k_l\G^{la}_\a+\xi^a_c\G^{kc}_\a\lb{SCcon}
\eea
which helps to construct the $SC$ covariant derivatives of superfields,
for instance,
\bea
\hat{D}^{kd}_\a W^{ab}=D^{kd}_\a W^{ab}-2\G^{kd}_\a
W^{ab}+ \G^{ka}_\a W^{bd}+\G^{kb}_\a W^{ad}
\eea
or its mirror image $\hat{D}^{kd}_\a L^{lj}$. The superfield
$W/L$ is the superconformal invariant.

\setcounter{equation}0
\section{$\cN=4$ tensor multiplets and hypermultiplets}

The three-dimensional left analytic hypermultiplet $q^{+a}$ has the free
action
\be
S^0_q=\frac12\int d\z^{-4}_Ldu\, q^+_a D^{++}q^{+a},\q \widetilde{q^{+a}}
=q^+_a.
\ee
The natural dimension of $q^{+a}$ is equal $\sfrac12$, and the corresponding
superconformal transformation contains the analytic parameter
\p{SC5}
\be
\de_{sc}q^{+a}=\la\, q^{+a},\qq \de_{sc}S^0_q=0.
\ee
This hypermultiplet interacts with the left $U_L(1)$ gauge prepotential
\bea
S(q^+,V^{++}_0)=\frac12\int d\z^{-4}_Ldu\, q^+_a [D^{++}q^{+a}+
(\tau_3)^a_b V^{++}_0q^{+b}]\lb{SqV}
\eea
where $\tau_3$ is the Pauli matrix. The mirror map $\cM S(q^+,V^{++}_0)=
S(Q^{(+)},A^{(++)}_0)$ yields the interaction of the right hypermultiplet
$Q^{(+)}_k$ with the right abelian prepotential $A^{(++)}_0$.

The dual  free $\om$-hypermultiplet can be described analogously
\be
S^0_\om=-\frac12\int d\z^{-4}_Ldu\, D^{++}\om D^{++}\om,\qq
\de_{sc}\om=\la\,\om.
\ee
In the gauge group $SU(N)$, we can use the adjoint representation
for the $\om$ superfield, then the hypermultiplet-gauge interaction
reads
\bea
&&S(\om,V)=-\frac12\int d\z^{-4}_Ldu\,\Tr (D^{++}\om+[V^{++},\om])^2.
\eea
The sum of this action and the $\cN=4$ Yang-Mills action
$S_N(g_3^2,V^{++})$ is invariant under the $\cN=8$ supersymmetry
transformations constructed by the analogy with the $D=4, \cN=4$
case \cite{GIOS}
\bea
\de V^{++}=g_3\hat\eps^{ka\a}u^+_k\te^+_{a\a}\om,\q
\de\om=-\frac{1}{2g_3}(D^+)^4(\hat\eps^{ka\a}u^-_k\te^-_{a\a}V^{--}),
\eea
where $\hat\eps^{ka\a}$ are spinor parameters.

The left tensor multiplet is described by the left analytic superfield
$L^{++}=u^+_ku^+_lL^{kl}(z)$
\bea
&&L^{++}
=-\frac14u^+_ku^+_l\int dv\, v^{(-)}_av^{(-)}_b D^{ka\a}D^{lb}_\a
A^{(++)}_0(z,v),\q D^{+a}_\a L^{++}=0,\lb{L0str}\\
&&D^{++}L^{++}=0,\q\de_{sc}L^{++}=2\la L^{++}.
\eea
The component representation of this superfield is
\bea
&&L^{++}=u^+_ku^+_l\La^{kl}-i\T^{++m}u^+_ku^-_l\pa^L_m\La^{kl}
-i\te^{+\a}_a u^+_k\rh^{ka}_\a+i\T^{++}_{ab}Y^{ab}\nn\\
&&+i(\te^{+b}\g^m\te^+_b)\ve_{mnp}\pa^nB^p+\T^{+3\a}_au^-_k
\pa_\ab\rh^{ka\bt}+(\te^+)^4u^-_ku^-_l\Box \La^{kl}.\lb{Lcomp}
\eea
The free action of the left tensor multiplet is equivalent to the
free action of the right gauge multiplet.

Now we consider the improved form of the left analytic tensor superfield
\bea
&&L^{++}=\g(c^{++}+l^{++}),\q c^{\pm\pm}=c^{kl}u^\pm_ku^\pm_l,\q
c^{++}c^{--}-(c^0)^2=1,
\nn\\
&&\de_{sc}l^{++}=2\la(l^{++}+c^{++})-2\la^{++}_Lc^0.
\eea
The  three-dimensional superconformal interaction of $l^{++}$ is similar
to the analogous  four-dimensional action \cite{GIOS}
\bea
&&\tilde{S}^L_0=-\frac{1}{\g}\int d\z^{-4}du(g^{++})^2,
\q g^{++}(l)=\frac{l^{++}}{1+\sqrt{1+l^{++}c^{--}}}.\lb{duSL}
\eea
We note that this action is a dual form of the action \p{SL0}.
The scalar part of the component action of the improved tensor multiplet
has the form
\bea
&&\tilde{K}_L=\int d^3x\frac{\sqrt{2}}{\La}\{\frac14\pa_m\La^{kl}
\pa^m\La_{kl}-\frac1{6\La^2}\La^{rs}\La^{kl}\pa_m\La_{kl}\pa^m\La_{rs}
+\frac14Y^{ab}Y_{ab}\}\lb{Lkinet}\\
&&\La=\sqrt{\La^{kl}\La_{kl}}=\sqrt{2}\g\sqrt{1+c^{kl}\la_{kl}
+\sfrac12\la^{kl}\la_{kl}},\q\La^{kl}=\g(c^{kl}+\la^{kl}).\nn
\eea
The similar component $D=4, \cN=2$ action was defined in \cite{WPV}.

As it was shown in \cite{GIOS}, the action of the improved tensor
multiplet is  dual to the action of the free hypermultiplet.
The alternative form of the action \p{duSL} contains unconstrained
superfield $l^{++}$ (or $g^{++}$) and the analytic Lagrange multiplier
$\om$
\bea
&&\tilde{S}^L_0=-\frac{1}{\g}\int d\z^{-4}_Ldu\left\{[g^{++}(l)]^2
-\om D^{++}l^{++}\right\}.
\eea
Using the algebraic equation $g^{++}(\om)=-(1+c^{--}D^{++}\om)^{-1}D^{++}\om$
we obtain the action $S(\om)$ which is equivalent to the free
hypermultiplet action. Thus the superconformal right gauge action \p{SL0}
is dual to the free action.

The link between the $CB$ and $LAB$ representations of the left
tensor multiplet is the formula
\bea
&&L^{-1}(l_{kl})=\frac{1}{\sqrt2\g}(1+c^{kl}l_{kl}+
\sfrac12l^{kl}l_{kl})^{-1/2}=\frac{1}{\sqrt2\g}\int du
[1+Z(l,u)]^{-3/2},\nn\\
&&Z(l,u)=l^{++}c^{--}=l^{kl}(z)c^{jn}u^+_ku^+_lu^-_ju^-_n.\lb{CBAB}
\eea
We note that both parts of this relation describe the same
$SU_L(2)$-invariant solution of the Laplace equation in variables $l_{kl}$:
$\D_l L^{-1}(l_{kl})=0$. This integral representation is based on the
formula
\be
\int du(l^{++}c^{--})^n=\frac{1}{2n+1}l^{(i_1k_1}l^{i_2k_2}\cdots l^{i_nk_n)}
c_{(i_1k_1}c_{i_2k_2}\cdots c_{i_nk_n)}
\ee
where brackets mean the total symmetrization of the $(l_{kl})$-polynomials
in $2n$ indices.

The superconformal interaction $S^W_0$ \p{S0} is equivalent to the  action
of the improved right tensor multiplet $w^{(++)}$
\bea
W^{(++)}=v^{(+)}_av^{(+)}_bW^{ab}=\vp(C^{(++)}+w^{(++)})
\eea
which is mirror to the left action \p{duSL}. The corresponding scalar terms
are
\bea
&&K_R=\int d^3x\frac{\sqrt{2}}{\phi}\left\{\frac14\pa_m\phi^{ab}\pa^m\phi_{ab}
-\frac1{6\phi^2}\phi^{cd}\phi^{ab}\pa_m\phi_{ab}\pa^m\phi_{cd}+
\frac14X^{kl}X_{kl}\right\}\lb{Rkinet}\\
&&\phi=\sqrt{\phi^{ab}\phi_{ab}}=\vp\sqrt{2+2C^{ab}\vp_{ab}+
\vp^{ab}\vp_{ab}}.\nn
\eea
The action of the improved right tensor multiplet $w^{(++)}$
is also dual to some free action.

\setcounter{equation}0
\section{Nonlinear $\cN=4$ gauge interactions}

Now we discuss the possible $\cN=4$ superfield  gauge interactions which
correspond to the higher degrees of the derivative terms $F_{mn}=\pa_mA_n
-\pa_nA_m,~F^B_{mn}=\pa_mB_n-\pa_nB_m,~\pa_m\phi^{ab}$ and $\pa_m\La^{kl}$
in the component actions. The nonlinear supersymmetric $D=4$
abelian gauge terms were analyzed in \cite{BIK},\cite{NL},\cite{BKT}.

We choose the nonlinear self-interaction terms  of the analytic  multiplet $L^{++}$
 in the following form:
\bea
&&S(L^{++})=\frac1{g_3^2}\int d\z^{-4}du\{(L^{++})^2[1-12c^2(D^+)^4(L^{--})^2]
\nn\\
&&
-24c^2L^{++}(D^+)^4[(L^{+-})^2L^{--}]\}+O(L^6),\lb{LNL}\\
&&L^{++}=u^+_ku^+_lL^{kl}(A^{(++)}_0),\q
L^{+-}=\sfrac12D^{--}L^{++},\q L^{--}=\sfrac12(D^{--})^2L^{++}\nn
\eea
where $c$ is some constant of the dimension -2. In components, these terms
describe the second and fourth degrees of $F^B_{mn}$
and $\pa_m\La^{kl}$.

By the analogy with the nonlinear realizations of the 4-dimensional
supersymmetries \cite{BIK} we can find the nonlinear $f$-transformation
of the superfield $L^{++}$
\bea
&&\de_f L^{++}=f^{++}[1+c^2(D^+)^4(L^{--})^2]+2c^2L^{++}(D^+)^4
(f^{--}L^{--})-4c^2(D^+)^4(f^{+-}L^{+-}L^{--})\nn\\
&&+4c^2(D^+)^4[f^{--}(L^{+-})^2]+O(L^4),\lb{ftrans}\\
&&f^{++}=a^{kl}u^+_ku^+_l+c^{-1}\te^{+\a}_a u^+_k\hat\eps^{ka}_\a,\q
f^{+-}=\sfrac12D^{--}f^{++},\q f^{--}=\sfrac12(D^{--})^2f^{++}\nn
\eea
where $a_{kl}$ and $\hat\eps^{ka}_\a$ are parameters of the bosonic and
fermionic translations. This transformation satisfies the condition
$D^{++}\de_f L^{++}=0$.

The nonlinear action \p{LNL} is invariant under the $f$-transformation
up to the third order in superfields. To prove this invariance we take into
account only two independent third-order structures in the variation
$\de_f S(L^{++})$
\be
f^{++}L^{++}(D^+)^4(L^{--})^2,\q (L^{++})^2(D^+)^4(f^{--}L^{--}).
\ee
The transformation \p{ftrans} describes the spontaneous breakdown of the
$D=3, \cN=8$ supersymmetry. The similar nonlinear action can be found for
the superfield $W^{(++)}$ in the right analytic superspace.

The fourth order nonlinear superfield terms can also be studied in the
full $D=3, \cN=4$ superspace
\bea
&&S_4=\int d^{11}z [A_1W^4+A_2L^4+A_3W^2L^2],\lb{NL4}\\
&&W^2=\vp^2(2+2C^{ab}w_{ab}+w^{ab}w_{ab}),\q
L^2=\g^2(2+2c^{kl}l_{kl}+l^{kl}l_{kl})\nn
\eea
where $A_1, A_2$ and $A_3$ are some constants.

Higher degrees of the gauge field strength arise from the superfield
terms with the spinor derivatives of $W^{ab}$, for instance,
\bea
&&S_6\sim \int d^{11}z W^4 A_\ab A^\ab,\nn\\
&&A_\ab=\ve_{kl}D^{ka}_{(\a}D^{lb}_{\bt)}W_{ab}\sim iF_\ab(x)+O(\te).
\lb{NL6}
\eea
All these nonlinear gauge interactions break down the superconformal
symmetry.

The superconformal $\cN=2, D=4$ nonlinear gauge interactions were
considered in \cite{BKT} where the chiral dilaton superfield was used.
We have two $\cN=4$ abelian dilaton-type superfields $W$ and $L$ so it
is not difficult to construct the  non-polynomial superconformal
generalization of the nonlinear terms $S_4$
\bea
S_{NL}=\int d^{11}z\left\{W^4L^{-5}+L^4W^{-5}\right\}.
\eea
We note that additional superconformal terms with the $SC$-invariant
combination $W/L$ could be added  to this interaction. Using the
connection \p{SCcon} we can construct the superconformal
generalizations of the derivative terms \p{NL6}.

We can also study the nonlinear superconformal interactions of the
nonabelian superfield strength \p{Wstr} with the abelian dilaton
superfields, for instance,
\bea
\int d^{11}z L^{-5}(\Tr\cW^{ab}\cW_{ab})^2.
\eea

\setcounter{equation}0
\section{ $\cN=4$ $BF$ interaction of the left and right
gauge multiplets}

We see that the left and right $\cN=4$ supermultiplets live in different
analytic superspaces, so it is extremely difficult to construct
interactions of these supermultiplets. Nevertheless, there is the simple
left-right gauge $BF$ interaction which was considered in the component
fields\cite{BG,KS} and also  in the biharmonic $\cN=4$ superspace \cite{Z6}.

The $LAB$ form of this $BF$ interaction reads
\bea
&&S^0_{BF}=\frac{i}2\bt\int d\z^{-4}_Ldu V^{++}_0L^{++}_0,
\eea
where $\bt$ is the coupling constant, $V^{++}_0$ is the left abelian
prepotential,  and $L^{++}$ is the scalar analytic
superfield strength  of
the right abelian pseudoscalar  prepotential
$A^{(++)}_0$ \p{L0str}.
This interaction  is manifestly superconformal
and preserves the $P$-parity and the $\cM$ symmetry
$V^{++}_0\leftrightarrow
A^{(++)}_0$.

The equations of motion for the abelian $BF$ model
\be
W^{ab}(V)=0,\q L^{++}(B)=0
\ee
have the pure gauge solutions only.

Using the superfield decompositions \p{WZ} and \p{Lcomp} we obtain
the component form of the abelian $\cN=4$ $BF$  action \cite{BG,KS}
\bea
S^0_{BF}=\bt\int d^3x(2\ve^{mnp}A_m\pa_nB_p -\sfrac12\phi^{ab}Y_{ab}
-\sfrac12\La^{ik}X_{ik}+2\rh^{ka}_\a\la^\a_{ka}).
\eea
The mirror symmetry $S^0_{BF}=\cM S^0_{BF}$ is evident in this
representation.

If we identify $SU_L(2)$ and $SU_R(2)$ indices of these fields, this
action can be treated as the difference of two abelian Chern-Simons
actions for two $\cN=3$ vector multiplets connected by the parity
transformation and the transformation of the fourth supersymmetry
\cite{BILPSZ}. The gauge $\cN=3$ prepotentials and superfield strengths
live in the same analytic superspace, and the fourth supersymmetry
transformation connect different gauge supermultiplets. We note that
the $\cN=4$ superfield generalization of the  Chern-Simons action
does not exist for the group $U(1)$, because the corresponding gauge
prepotentials and superfield strengths are defined in different superspaces.

One can add the minimal interactions of the left hypermultiplets
$q^+_a$ \p{SqV} and the right hypermultiplets $Q^{(+)}_k=\cM q^+_a$ to the
$U_L(1)\times U_R(1)$ $BF$ action
\be
S(q,Q,V_0,A_0)=S^0_{BF}+S(q^+,V^{++}_0)+S(Q^{(+)},A^{(++)}_0).
\ee
We note that this model has the manifest $\cN=4$ symmetry even if it be
reformulated in the $\cN=3$ superspace. In this formalism, generators of
the linear transformations of the fourth supersymmetry have the opposite
sign on the left and right $\cN=3$ analytic superfields: $Q^4_\a=
\pm D^0_\a$. More complex nonlinear transformations of the higher
supersymmetries are defined in \cite{BILPSZ} for the nonabelian $\cN=3$
superfields, the algebra of these transformations closes only on the
equations of motion.

The abelian $BF$ term can be treated as the nontrivial interaction
in the following superconformal composite action:
\be
S^\bt(V^{++}_0,A^{(++)}_0)=S^W_0+S^L_0+S^0_{BF}\lb{LRgauge}
\ee
where the first two terms describe the left and right gauge superfields.
It is evident that this interaction possesses the discrete
$\cM$ symmetry.

The scalar part of this action contains the kinetic term for the field
$\La^{kl}$ \p{Lkinet}, the analogous kinetic term for the field
$\Phi^{ab}$ \p{Rkinet} and the mixed potential term
\be
P=-\frac{\sqrt{2}\bt^2}{4}\int d^3x[\phi^{ab}\phi_{ab}\sqrt{\La^{kl}\La_{kl}}
+\La^{kl}\La_{kl}\sqrt{\phi^{ab}\phi_{ab}}]
\ee
which arises from terms with the auxiliary fields $X^{kl}$
and $Y^{ab}$. Thus, the superconformal action $S^\bt(V^{++}_0,A^{(++)}_0)$
describes the nontrivial interaction of the abelian left and right $\cN=4$
gauge multiplets.

Using the transformation of the vector and pseudovector gauge fields in
\p{LRgauge}
\bea
&&A_m=\sqrt{\frac{\vp}{2}}(A^+_m+A^-_m),\q
B_m=\sqrt{\frac{\g}{2}}(A^+_m-A^-_m)
\eea
we obtain the sum of two quadratic gauge actions
\bea
&&S(A^+_m,A^-_m)=\frac12\int d^3x[A^+_mK^{mn}_+A^+_n+A^-_mK^{mn}_-A^-_n],
\nn\\
&&K^{mn}_\pm=\eta^{mn}\Box-\pa^m\pa^n\pm\mu\ve^{mpn}\pa_p
\eea
where $\mu=\bt\sqrt{\vp\g}$ is the parameter of the gauge-invariant mass
terms.

It is easy to build the non-abelian version of the $BF$ theory in the
full $\cN=4$ superspace using the superfield $\cW^{ab}$ \p{Wstr} and the
Lagrange multiplier $B_{ab}$, however, we do not know interactions of
this additional superfield $B_{ab}$. We also cannot construct interactions
of the left $\cN=4$ non-abelian gauge superfield with the right
non-abelian gauge superfield.

\section{Conclusions}
We analyzed the superconformal interactions of $D=3, \cN=4$
superfields in different representations. The left and right
$\cN=4$ harmonic superspace are treated as two mirror analogs
of the $D=4, \cN=2$ harmonic superspace. The interaction of
the left gauge and left hypermultiplet superfields are natural,
and the mirror picture connects right $\cN=4$ superfields, while
it is difficult to construct interactions of the left and right
supermultiplets. We consider the dilaton superfield $W$ using
the abelian gauge superfield strength $W^{ab}$ and the mirror dilaton
superfield $L$ constructed from the right abelian gauge superfield.
$W$ and/or $L$ play the role of the dynamic coupling constants in the
superconformal gauge theory. These dilatons allow us to construct
the improved superconformal versions of the abelian and nonabelian
$\cN=4$ gauge theories. The  left $\cN=4$ superconformal abelian gauge
model $S^W_0$ is dual to the free hypermultiplet theory, this property is
preserved in the mirror model $S^L_0$.

The  $BF$ interaction in the $N=4$ superspace is considered as the
 analog of the Chern-Simons action for the group $U_L(1)\times U_R(1)$.
The left and right hypermultiplets interact with the corresponding
$\cN=4$ gauge superfields in this theory. We propose the combining
superconformal interaction of the abelian left and right gauge superfields
including the improved left and right gauge actions $S^W_0+S^L_0$
and  the mixing $BF$ term. This term yields the nontrivial
interaction of the scalar and vector fields from two mirror
supermultiplets.

On the non-superconformal level, we find the nonlinear realization
of the $\cN=8$ supersymmetry connecting different nonlinear gauge terms
in the effective action. Using the superconformal covariant derivatives
of the superfield strengths we obtain the higher nonlinear superconformal
interactions of the left and right gauge superfields.

The author is grateful to E.A. Ivanov for interesting discussions. This
work was partially supported by the grants RFBR N 09-02-01209,
09-01-93107-CNRS and  08-02-90490-Ukr, DFG 436 RUS 113/669/0-4R,
INTAS 05-10000008-7928 and by the Heisenberg-Landau programme.

\renewcommand\theequation{A.\arabic{equation}}
\setcounter{equation}0
\section*{Appendix }

{\bf 1. Central basis of  $D=3, \cN=4$  superspace}\\

We consider the coordinates of the $D=3, \cN=4$ superspace in the
central basis $(CB)$ and the corresponding integral measure \cite{Z3,Z5}:
\bea
&&z=(x^m, \te^\a_{ka}),\q d^{11}z=d^3x\,d^8\te,
\eea
where $i$ and $a$ are the two-component indices of the automorphism
groups $SU_L(2)$ and $SU_R(2)$, respectively, $\a$ is the two-component
index of the $SL(2,R)$ group and $m=0, 1, 2$ is the 3D vector index. The
three-dimensional $\gamma$ matrices satisfy the relations
\bea
&&(\gamma_m)_\alpha^\rho(\gamma_n)_\rho^\beta=-(\gamma_m)_{\alpha\rho}
(\gamma_n)^{\rho\beta}
=-\eta_{mn}\delta^\beta_\alpha+\varepsilon_{mnp}(\gamma^p)^\beta_\alpha,
\eea
where $\eta_{mn}=\mbox{diag}(1,-1,-1)$ and $\ve_{mnp}$ is the antisymmetric
symbol.

We use the $P$-parity transformation
\be
Px^{0,2}=x^{0,2},\q Px^1=-x^1,\q P\te^\a_{ka}=(\g_1)^\a_\bt \te^\bt_{ka}.
\ee

The mirror map $\cM$ interchanges values of the isospinor indices
of the groups $SU_L(2)$ and $SU_R(2)$, for instance, $\cM\theta^\a_{12}
=\theta^\a_{21}$, and connects the representations $(l,r)$ and $(r,l)$.

The $\cN=4$ spinor derivatives have the form
\bea
&&D^{ka}_\a=\pa^{ka}_\a+i\te^{ka\bt}\pa_\ab,\q \pa_\ab=(\g^m)_\ab\pa_m.
\eea

{\bf 2. Left $D=3, \cN=4$ harmonic superspace}\\

The left $D=3, \cN=4$ harmonic superspace was considered in \cite{Z3,Z5}.
It uses the left $SU(2)_L/U(1)$ harmonics $u^\pm_k$ and the corresponding
left analytic basis $(LAB)$
\bea
&&\z_L=(x^m_L, \te^{+\a}_a),\q \te^{-\a}_a,\nn\\
&&x^m_L=x^m+i(\g^m)_\ab\te^{+a\a}\te^{-\bt}_a,\q \te^{\pm\a}_a=
u^\pm_k\te^{k\a}_a.\lb{LAB}
\eea
We define the  special conjugation in this basis
\bea
\widetilde{u^\pm_k}=u^{\pm k},\q \widetilde{x^m_L}=x^m_L,\q
\widetilde{\te^{\pm\a}_a}=\te^{\pm a\a}.\lb{spec}
\eea

The $LAB$ spinor and harmonic derivatives are
\bea
&&D^{+a}_\a=\pa^{+a}_\a,\q D^{-a}_\a=-\pa^{-a}_\a+2i\te^{-a\bt}\pa^L_\ab,
\lb{Lspin}\\
&&D^{++}=\pa^{++}-i\te^{+}_a\g^m\te^{+a}\pa^L_m+\te^{+\a}_a\pa^{+a}_\a,\nn\\
&&D^{--}=\pa^{--}-i\te^{-}_a\g^m\te^{-a}\pa^L_m+\te^{-\a}_a\pa^{-a}_\a.
\lb{LAB2}
\eea
The simple combinations of the spinor derivatives are
\bea
&&D^{\pm\pm ab}=D^{\pm a\a}D^{\pm b}_\a,\q D^{\pm\pm\ab}=
D^{\pm a}_\a D^{\pm}_{a\bt},\nn\\
&&(D^\pm)^4=\frac{1}{48}D^{\pm\pm ab}D^{\pm\pm}_{ab}
=\frac1{16}(D^{\pm1})^2(D^{\pm2})^2.
\eea
They satisfy the relations
\bea
&&D^{++ab}D^{++\ab}=0,\\
&&(D^+)^4(D^{--})^2(D^+)^4=2\pa^m\pa_mD^{(+4)}=2\Box (D^+)^4,\nn\\
&&(D^+)^4(D^{--})^3(D^+)^4=[\sfrac{3i}2D^{-b}_\a D^{-}_{b\bt}\pa^\ab
+6D^{--}\Box] (D^+)^4=[\sfrac{3i}2\pa^{-b}_\a \pa^{-}_{b\bt}\pa^\ab
+6\pa^{--}\Box] (D^+)^4,\nn\\
&&(D^+)^4(D^{--})^4(D^+)^4=[\sfrac12D^{--}_{ab}D^{--ab}+6iD^{--}D^{-a\a}
D^{-\bt}_a\pa_\ab+12(D^{--})^2]\Box(D^+)^4\nn\\
&&=[24(\pa^-)^4+6i\pa^{--}\pa^{-a\a} \pa^{-\bt}_a\pa_\ab
+12(\pa^{--})^2]\Box(D^+)^4.\lb{A3}
\eea

The analytic integral measure in $LAB$ is
\be
d\z^{-4}_Ldu=d^4x_Ldu(D^-)^4\lb{Lint}
\ee
where $du$ describes the integration on $SU_L(2)/U(1)$.

We use the basic combinations of the left analytic spinor
coordinates $\te^{\pm\a}_a$
\bea
&&\T^{\pm\pm}_{ab}=\te^{\pm\bt}_a\te^\pm_{b\bt},\q \T^{\pm\pm}_{\ab}=
(\g^m)_\ab\T^{\pm\pm}_m=\te^{\pm a}_\a\te^\pm_{a\bt},\nn\\
&&\T^{\pm3\a}_a=\sfrac23\te^{\pm b\a}\T^{\pm\pm}_{ab},\q (\te^\pm)^4=
(\te^\pm_1)^2(\te^\pm_2)^2.
\eea
They satisfy the simple identities of the Grassmann algebra
\bea
&&\T^{\pm\pm}_{ab}\T^{\pm\pm}_{cd}=\sfrac12(\ve_{ac}\ve_{bd}+\ve_{ad}\ve_{bc})
(\te^\pm)^4,\nn\\
&&\T^{\pm\pm}_{\ab}\T^{\pm\pm}_{\g\rh}=-\sfrac12(\ve_{\a\g}\ve_{\bt\rh}
+\ve_{\a\rh}\ve_{\bt\g})(\te^\pm)^4,\nn\\
&&\te^{\pm\a}_c\T^{\pm\pm}_{ab}=\sfrac12\ve_{ca}\T^{\pm3\a}_b+\sfrac12
\ve_{cb}\T^{\pm3\a}_a.
\eea
\\

{\bf 3. Right $\cN=4$ harmonic superspace}\\

We denote the mirror $SU_R(2)/U(1)$ harmonics as $v^{(\pm)}_a=\cM u^\pm_k$
and the coordinates of the right analytic basis $(RAB)$ as
\bea
&&\z_R=\cM\z_L=(x^m_R, \te^{(+)\a}_k),\q \te^{(-)\a}_k,\nn\\
&&x^m_R=x^m+i(\g^m)_\ab\te^{(+)k\a}\te^{(-)\bt}_k,\q
\te^{(\pm)\a}_k=v^{(\pm)}_a\te^{a\a}_k.\lb{RAB}
\eea
The special conjugation in $RAB$ is analogous to the
corresponding conjugation in $LAB$ \p{spec}.

The spinor and harmonic derivatives in the $RA$ basis can be obtained
by the mirror map from \p{LAB2}, for instance,
\bea
&&D^{(+)k}_\a=\pa^{(+)k}_\a,\q D^{(-)k}_\a=-\pa^{(-)k}_\a+2i\te^{(-)k\bt}
\pa^R_\ab,\nn\\
&&D^{(++)}=\pa^{(++)}-i\te^{(+)}_k\g^m\te^{(+)k}\pa^R_m+\te^{(+)\a}_k
\pa^{(+)k}_\a,\lb{RAB2}
\eea
where the  partial derivatives act on the corresponding right
coordinates. The right analytic superfields $\Phi(\z_R,v)$ describe right
$\cN=4$ supermultiplets.

\end{document}